\documentclass[twocolumn,showpacs,preprintnumbers,amsmath,amssymb,superscriptaddress]{revtex4}
\usepackage[toc,page]{appendix}
\usepackage{graphicx}
\usepackage{dcolumn}
\usepackage{bm}
\usepackage{color}
\usepackage{amsbsy}
\usepackage[hidelinks]{hyperref}

\begin{document}

\title{Sub-Poissonian atom-number distributions by means of Rydberg dressing and electromagnetically induced transparency}

\author{Daniel Cano}

\affiliation{ICFO-Institut  de  Ci\`{e}ncies  Fot\`{o}niques,  The  Barcelona  Institute of  Science  and  Technology,  08860  Castelldefels (Barcelona),  Spain}

\begin{abstract}
A method is proposed to produce atomic ensembles with sub-Poissonian atom number distributions. The method consists of removing the excess atoms using the interatomic interactions induced by Rydberg dressing. The selective removal of atoms occurs via spontaneous decay into untrapped states using an electromagnetically induced transparency scheme. Ensembles with the desired number of atoms can be produced almost deterministically. Numerical simulations predict a strong reduction of the atom number fluctuations, with the variance twenty times less than the Poisson noise level (the predicted Fano factor is $F \simeq 0.05$). Strikingly, the method is suitable for both fermions and bosons. It solves the problem of the atom-number fluctuations in bosons, whose weak interactions have usually been an obstacle to controlling the number of atoms.
\end{abstract}

\maketitle

\section{Introduction} \label{sec:introduction}

Ensembles of a few ultracold atoms are basic tools for quantum information and precision measurements \cite{Saffman:10,Pezze:18}. They can be used as model systems to investigate few-body quantum phenomena like tunneling \cite{Hensinger:01,Zollner:08,Rontani:12,Zurn:12}, interactions \cite{Stecher:07,Weiner:99,Blume:12,Cano:12}, Efimov states \cite{Efimov:70,Kraemer:06}, and quantum correlations \cite{Sorensen:03,Greiner:05,Koutentakis:19}. To fully exploit the potential of few-atom systems in these applications, the number of atoms $N$ must be controlled with high precision. In general, the precise control of $N$ is not trivial because the production of ultracold atoms is affected by atom number fluctuations caused by Poisson statistics and technical noise. Reducing, or even suppressing, atom number fluctuations is a prerequisite for quantum computation based on atomic ensembles \cite{Lukin:01,Brion:07,Muller:09,Su:20,Zheng:20a} and is the first step in many entanglement protocols to achieve measurement precision beyond the shot noise \cite{Wineland:94,Saffman:09,Toth:14,Cano:14,Opatrny:16,Zheng:20b}.

To produce ensembles of a few atoms (up to 100), the usual experimental procedure consists of removing the excess atoms by means of interatomic interactions. The atoms whose interatomic interaction energy surpasses the atom trap depth are eventually lost, resulting in an atomic ensemble with sub-Poissonian atom number distribution. This procedure relies on the precise experimental control of the interatomic interactions and the atom trap parameters \cite{Campo:08,Pons:09,Sokolovski:11}. Different kinds of interactions have been used, including dispersive $s$-wave interactions \cite{Chuu:05}, Pauli blockade in fermions \cite{Serwane:11,Wenz:13}, three-body inelastic collisions \cite{Whitlock:10,Itah:10}, and light-assisted two-body collisions \cite{Schlosser:01,Nelson:07,Sortais:12}. All these studies show important differences between fermions and bosons. Fermions provide a greater capacity to produce the desired $N$ in a deterministic way thanks to the strong repulsive interactions originating from the Pauli principle. Ensembles of up to ten ground-state fermions have been achieved with high fidelity \cite{Serwane:11}. In contrast, bosons offer less favorable conditions due to their weaker interactions. Besides the experimental efforts, the fluctuation factor (Fano factor) achieved with bosons is typically $F \simeq 0.5$, which is half that of the Poisson distribution. Reducing the Fano factor to lower values remains a desired objective.

This paper presents an alternative method that is suitable for both fermions and bosons. To produce sub-Poissonian distributions, the method uses the interatomic interactions induced by the so-called Rydberg dressing \cite{Balewski:14,Gil:14}. This consists of slightly mixing a ground state with a highly-excited Rydberg state through non-resonant laser coupling. In this way, the ground state partially acquires the strong interaction properties of the Rydberg states \cite{Low:12,Singer:04,Tong:04,Jaksch:00}. As shown in this paper, these interactions enable the removal of the excess atoms and the production of strongly sub-Poissonian distributions. As opposed to previous works, atom losses are induced by spontaneous decay into untrapped states in an electromagnetically induced transparency scheme. Numerical calculations predict Fano factors as low as $F \simeq 0.05$. Starting with a random number of atoms in a magnetic trap, the proposed method induces spontaneous atom loss until the target number of atoms is reached.

\section{Method to generate sub-Poissonian distributions} \label{sec:methods}

\begin{figure}
\centerline{\scalebox{0.24}{\includegraphics{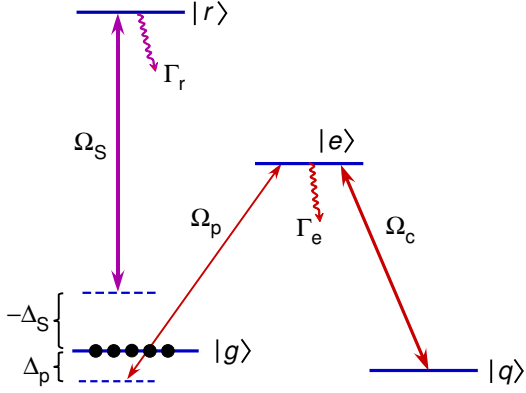}}}
\caption{Energy level scheme. The initial ground state $|g\rangle$ is slightly mixed with a Rydberg state $|r\rangle$ by an off-resonant laser field with Rabi frequency $\Omega_{\rm S}$ and detuning $\Delta_{\rm S}$. This field is used to induce interatomic interactions between the Rydberg-dressed ground-state atoms. The ground state $|g\rangle$ is coupled to state $|e\rangle$ using a laser field with Rabi frequency $\Omega_{\rm p}$ and detuning $\Delta_{\rm p}$. At the same time, $|e\rangle$ is resonantly coupled to a second Zeeman ground state $|q\rangle$ with Rabi frequency $\Omega_{\rm c} \gg \Omega_{\rm p}$, thus creating a $\Lambda$-type electromagnetically induced transparency configuration. The decay rates of $|e\rangle$ and $|r\rangle$ are $\Gamma_e$ and $\Gamma_r$, respectively. The black dots represent a possible initial population in $|g\rangle$.} \label{fig:AtomicLevels}
\end{figure}

The method uses the atomic level structure of Fig. \ref{fig:AtomicLevels}. All atoms are initially prepared in the same Zeeman ground state, $|g\rangle$. Atoms interact with each other by means of Rydberg dressing. For this, the state $|g\rangle$ is slightly mixed with a Rydberg state $|r\rangle$ using an off-resonant laser field with Rabi frequency $\Omega_{\rm S}$ and detuning $\Delta_{\rm S} \gg \Omega_{\rm S}$. The AC Stark shift of the collective ground state $|g^N\rangle \equiv |g_1,g_2,...,g_N\rangle$ can be approximated by (Appendix \ref{sec:StarkShift}),
\begin{equation}
{\Delta}_{\rm AC}(N) \simeq \frac{\Omega_{\rm S}^2}{4 \Delta_{\rm S}} N - \frac{\Omega_{\rm S}^4}{16 \Delta_{\rm S}^3} N^2,
\label{Eq:StarkShift}
\end{equation}
where the nonlinear term in $N$ originates from the Rydberg blockade mechanism, i.e. the suppression of multiple Rydberg excitations due to the strong dipolar interactions. This term represents the interatomic interaction energy that enables the controlled removal of atoms.

The method consists of removing the excess atoms until the ensemble only contains the target number of atoms $N_{\rm T}$. The goal is that the magnetic trap contains $N_{\rm T}$ atoms at the end of the process. The target state is $|g^{N_{\rm T}}\rangle$. Atoms are removed by means of spontaneous decay from the short-lived state $|e\rangle$ into magnetically untrapped states. The precise control of the spontaneous decay rate is achieved by means of electromagnetically induced transparency (EIT). For this, a weak laser beam couples $|g\rangle$ to $|e\rangle$ with Rabi frequency $\Omega_{\rm p}$ and detuning $\Delta_{\rm p}$ while another laser beam resonantly couples $|e\rangle$ to a second Zeeman ground state $|q\rangle$ with Rabi frequency $\Omega_{\rm c} \gg \Omega_{\rm p}$. The method requires that the atom loss rate $\Gamma_N$ is close to zero for $N=N_{\rm T}$ and increases as $|N-N_{\rm T}|$ increases. This condition is fulfilled when the field $\Omega_{\rm p}$ is resonant with the transition $|g^{N_{\rm T}}\rangle \leftrightarrow |g^{N_{\rm T}-1} e\rangle$ (this is the standard notation of the collective symmetric states \cite{Lukin:01}; see below). This occurs for
\begin{eqnarray}
\Delta_{\rm p} &=& {\Delta}_{\rm AC}(N_{\rm T}-1)-{\Delta}_{\rm AC}(N_{\rm T}) \nonumber \\ &\simeq& -\frac{\Omega_{\rm S}^2}{4\Delta_{\rm S}} + (2N_{\rm T}-1)\delta, \label{Eq_DeltaP}
\end{eqnarray}
where
\begin{equation}
\delta = \frac{\Omega_{\rm S}^4}{16 \Delta_{\rm S}^3} \label{Eq_CharacteristicShift}
\end{equation}
is the characteristic energy shift of the interatomic interaction. The detuning $\Delta_{\rm p}$ has higher-order terms in $N$ that are not written in Eq. \ref{Eq_DeltaP}. It is not worth writing those terms here because their relative value is small. Nonetheless, higher-order terms will be used in the calculations for high numerical precision.

To verify that the EIT scheme enables the removal of excess atoms, we need to calculate the atom loss rate $\Gamma_N$ as a function of $N$. For this, we numerically solve the Schr\"{o}dinger equation,
\begin{equation}
\frac{d}{dt} |\psi_N (t)\rangle = -\frac{i}{\hbar}{\cal H}_N(t) |\psi_N (t) \rangle, \label{Eq_Schroedinger}
\end{equation}
where ${\cal H}_N(t)$ is the Hamiltonian of the light-atom coupling for an ensemble with $N$ atoms. The Hamiltonian in the interaction picture is given by (within the rotating-wave approximation),
\begin{eqnarray}
\mathcal{H}_N = &-& \frac{\hbar}{2} \left( \Omega_{\rm p} e^{-{\rm i} \left( \Delta_{\rm p} + \Delta_{\rm l} \right) t} \hat{\sigma}_{eg} + \Omega_{\rm c} \hat{\sigma}_{eq} \right. \nonumber \\ &+& \left. \Omega_{\rm S} e^{-{\rm i} \Delta_{\rm S} t} \hat{\sigma}_{rg} + {\rm h.c.} \right)  \nonumber \\ &-&  {\rm i} \hbar \frac{\Gamma_e}{2} \hat{\sigma}_{ee} - {\rm i} \hbar \frac{\Gamma_r}{2} \hat{\sigma}_{rr}, \label{Eq_Hamiltonian}
\end{eqnarray}
where $\hat{\sigma}_{\mu\nu}=\sum_{j=1}^{N} \sigma_{\mu\nu}^{(j)}$ are the collective symmetric operators and $\sigma_{\mu\nu}^{(j)} = |\mu_j \rangle \langle \nu_j |$ are the atomic transition operators of atom $j$, with $\mu,\nu=g,q,e,r$. The detuning $\Delta_{\rm l}$ is added to the field $\Omega_{\rm p}$ in order to simulate an unintended frequency shift caused by possible experimental errors. In all numerical simulations, we use the decay rate of state $|5P\rangle$ of rubidium, $\Gamma_e = 2 \pi \times 6$ MHz \cite{Steck:19}, and the decay rate of the Rydberg state $|70P\rangle$, $\Gamma_r = 2 \pi \times 100$ Hz \cite{Beterov:09} (without black-body radiation \cite{Cano:08,Cano:11,Jessen:13}). Equation \ref{Eq_Schroedinger} is solved using the basis of symmetric states, \begin{equation}
|g^{\alpha} q^{\beta} e^{\gamma} r^\eta \rangle = \sqrt{\frac{\alpha! \beta! \gamma!}{N!}} \sum_k {\rm P_k} \{ |g\rangle^{\alpha} |q\rangle^{\beta} |e\rangle^{\gamma} |r\rangle^{\eta} \},
\end{equation}
where $\alpha+\beta+\gamma+\eta=N$, $\eta=0,1$, and ${\rm P_k}\{ \cdot \}$ denotes the complete set of the $N!/(\alpha! \beta! \gamma!)$ possible permutations of the single-atom states.

\begin{figure} 
\centerline{\scalebox{0.44}{\includegraphics{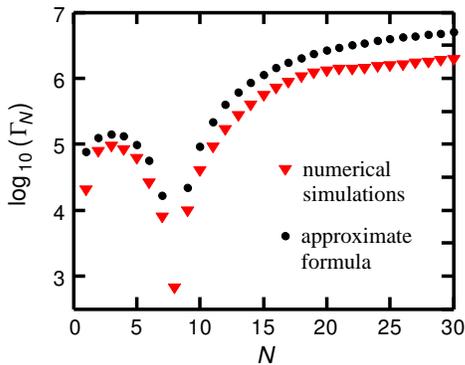}}}
\caption{Atom loss rate as a function of the number of atoms in the ensemble. The target number is $N_{\rm T}=8$ and the interaction energy shift is $\delta = 2 \pi \times 20$ kHz (see Eq. \ref{Eq_CharacteristicShift}). The laser parameters are $\Omega_{\rm c} = 2 \pi \times 2$ MHz, $\Omega_{\rm p} = 2 \pi \times 400$ kHz, and $\Delta_{\rm S} = 2 \pi \times 300$ MHz (from these parameters we obtain $\Delta_{\rm p} \simeq - 2 \pi \times 2.2$ MHz and $\Omega_{\rm S} \simeq 2 \pi \times 54$ MHz, using Eqs. \ref{Eq_DeltaP} and \ref{Eq_CharacteristicShift}). The results are obtained from numerical simulations using the multi-atom Hamiltonian of Eq. \ref{Eq_Hamiltonian} (red triangles) and from the approximate formula described in Appendix \ref{sec:BlochEquations} (black circles).} \label{fig:AtomLossRate}
\end{figure}

The atom loss rate is calculated as $\Gamma_N = \Gamma_e P_e$, where $P_e = \sum_{j=1}^N | \langle e_j | \psi_N \rangle|^2$ is the total population in $|e\rangle$, and $| \psi_N \rangle$ is the steady-state wavefunction of the ensemble with $N$ atoms. We have checked that spontaneous decay from $|r\rangle$ is negligible in comparison with spontaneous decay from $|e\rangle$. In fact, the Rydberg population, $P_{r} \simeq (N \Omega_{\rm S}^2)/(4 \Delta_{\rm S}^2)$, is only $\sim 7$\% of the total population for $N=N_{\rm T}$ in the numerical examples of this study. Figure \ref{fig:AtomLossRate} shows the atom loss rate $\Gamma_N$ for the target number $N_{\rm T}=8$. There is a sharp minimum at $N_{\rm T}$, where the EIT effects are maximum. Thus, for any initial atom number higher than $N_{\rm T}$, atom losses occur quickly until the number of atoms stabilizes for $N=N_{\rm T}$, resulting in a sub-Poissonian atom number distribution. Figure \ref{fig:AtomLossRate} compares the numerical results obtained using the multi-atom Hamiltonian ${\cal H}_N$ (Eq. \ref{Eq_Hamiltonian}) with the results obtained using an approximate formula derived from the optical Bloch equations of a three-level atom (see Appendix \ref{sec:BlochEquations}). Only the numerical solution is accurate, as only this is calculated considering all collective states. Nonetheless, the approximate formula from the optical Bloch equations follows the same tendency as the numerical solution, and we use it to verify the underlying EIT origin of the atom-light coupling.

\section{Selective removal of excess atoms} \label{sec:trajectories}

Once we know the atom loss rate $\Gamma_N$ as a function of $N$, we can simulate the atom loss dynamics leading to sub-Poissonian distributions. The simulations assume that the initial number of atoms is unknown and follows Poissonian statistics. The numerical simulations are carried out using the stochastic method of quantum trajectories \cite{Dalibard:92,Lambropoulos:06}. We simulate a high number of quantum trajectories and calculate the mean value $\overline{N}$ and the variance $\sigma^2 \left( N \right)$ of the number of atoms. Each quantum trajectory represents a thought experiment in which the initial number of atoms $N_0$ is randomly chosen according to the initial Poissonian distribution. To consider possible errors of the laser frequency locking system, the simulations include a random shift $\Delta_{\rm l}$ in the detuning of the laser field $\Omega_{\rm p}$ (see Eq. \ref{Eq_Hamiltonian}). For each quantum trajectory, this frequency shift is chosen from normally distributed random numbers with the standard deviation $\sigma^2 \left( \Delta_{\rm l} \right)$. In this way, for each quantum trajectory, we calculate the series of times $\{t_{N_0},t_{N_0-1},t_{N_0-2},...\}$ at which the ensemble loses one atom,
\begin{equation}
N_0 \hspace{1mm} \xrightarrow{\hspace{0.5mm} \textit{\normalsize t}_{\textit{\scriptsize N}_0}\hspace{0.5mm}} \hspace{1mm} N_0-1 \hspace{1mm} \xrightarrow{\hspace{0.5mm} \textit{\normalsize t}_{\textit{\scriptsize N$_0-1$}}\hspace{1mm}} \hspace{0.5mm} N_0-2 \hspace{1mm} \xrightarrow{\hspace{0.5mm} \textit{\normalsize t}_{\textit{\scriptsize N$_0-2$}}\hspace{0.5mm}} \hspace{1mm} \cdots
\end{equation}
This series of times is obtained using the condition that spontaneous decay occurs when the squared norm of the wavefunction has decreased to $|\langle \psi_N | \psi_N \rangle |^2 = r_N$,  where $\{r_N\}_{N=1,...,N_0}$ is a series of random numbers between 0 and 1 \cite{Dalibard:92,Lambropoulos:06}. This condition can be rewritten as $\exp{\left[-(t_{N-1}-t_N) \Gamma_N \right]}=r_N$. From this expression, we can easily calculate the time interval between two events, $t_{N-1}-t_N = - \log{r_N} / \Gamma_N$, thus obtaining the series of times $\{t_{N_0},t_{N_0-1},t_{N_0-2},...\}$ of the spontaneous decay events of the quantum trajectory.

Spontaneous decay leads to the selective removal of atoms through optical pumping into magnetically untrapped states. For this purpose, the states $|g\rangle$, $|q\rangle$ and $|e\rangle$ must be conveniently chosen. One possibility could be, for example, the rubidium states $|g\rangle \equiv |5S_{1/2}, F=2, m_F=1 \rangle$, $|q\rangle \equiv |5S_{1/2}, F=1, m_F=-1 \rangle$, and $|e\rangle \equiv |5P_{3/2}, F=1, m_F=0 \rangle$. The probability that an atom decays from $|e\rangle$ into a ground state other than $|g\rangle$ is 95 \% of the total decay probability \cite{Steck:19}. In this case, the atom is optically pumped outside the basis of states, and it cannot be part of the collective dark state of the EIT scheme of Fig. \ref{fig:AtomicLevels}. This is guaranteed by the offset magnetic field of the trap, which breaks the degeneracy of the ground state with Zeeman shifts typically of the order of a few MHz \cite{Fortagh:07}. The atom that decays ends up in a magnetically untrapped ground state like $|5S_{1/2}, F=1, m_F=1 \rangle$ or $|5S_{1/2}, F=2, m_F=-1 \rangle$, which does not interact with the laser fields $\Omega_{\rm p}$ and $\Omega_{\rm c}$, and which eventually leads to the loss of the atom.

\section{Results} \label{sec:results}

\begin{figure}
\centerline{\scalebox{0.44}{\includegraphics{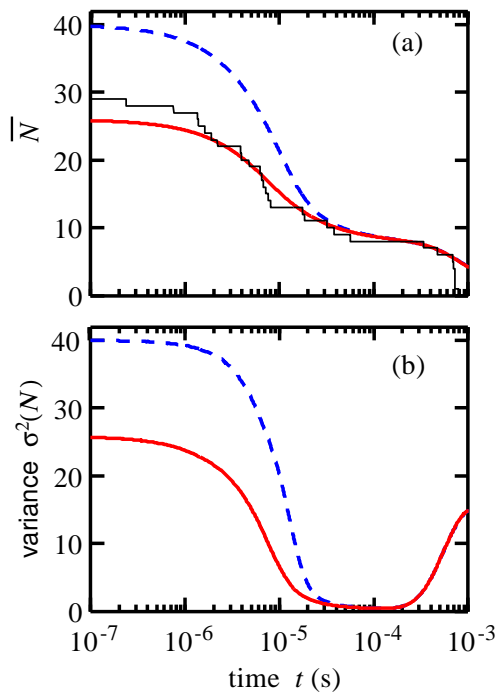}}}
\caption{Evolution of the number of atoms for the target number $N_{\rm T}=8$. The interaction energy shift is $\delta = 2 \pi \times 20$ kHz. The laser parameters are the same as those in Fig. \ref{fig:AtomLossRate}. Two initial Poissonian distributions are considered, with mean numbers $\overline{N_0}=$26 (red solid line) and $\overline{N_0}=$40 (blue dotted line). Each curve is obtained by averaging $2 \times 10^4$ simulated quantum trajectories. The variance of the random frequency noise is $\sigma^2 \left( \Delta_{\rm l} \right) = 2 \pi \times 6$ kHz. (a) Mean number of atoms as a function of time. The black solid curve shows one of the individual quantum trajectories with $N_0=29$. (b) Variance of the number of atoms as a function of time. The variance is strongly sub-Poissonian when $\overline{N}\simeq N_{\rm T}$.} \label{fig:Evolution}
\end{figure}

\begin{figure}
\centerline{\scalebox{0.44}{\includegraphics{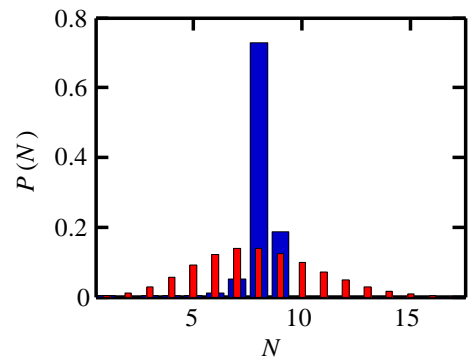}}}
\caption{Probability of finding $N$ atoms (wide blue bars) with the same parameters as in Fig. \ref{fig:AtomLossRate} at the time when $N=N_{\rm T}$ ($t \simeq 200 \mu$s). The narrow red bars represent the Poissonian distribution, for comparison.} \label{fig:Histogram}
\end{figure}

Figure \ref{fig:Evolution} shows the mean number of atoms $\overline{N}$ and the variance $\sigma^2(N)$ as a function of time in two simulated cases. The target number is $N_T=8$ in both cases, and the initial mean numbers are $\overline{N_0}=$ 26 and 40, respectively. Each curve is obtained from twenty thousand quantum trajectories (this is enough to converge). The mean number $\overline{N}$ decreases very quickly in the beginning of the process due to the high decay rates when $|N-N_T|$ is large (see Fig. \ref{fig:AtomLossRate}). The mean number stabilizes when it reaches the target number $N_T$. This happens at $t \simeq 200 \mu$s. At this moment, the laser fields must be turned off adiabatically, avoiding any remaining population in states $|q\rangle$ or $|r\rangle$. This is accomplished by turning off the laser fields in order, first $\Omega_{\rm p}$, then $\Omega_{\rm c}$, and finally $\Omega_{\rm S}$ \cite{Fleischhauer:05}. The atom number distribution is clearly sub-Poissonian. The variance has values as low as $\sigma^2(N) \simeq 0.4$. This corresponds to a Fano factor of $F=\sigma^2(N)/\overline{N} \simeq 0.05$, which is twenty times less than that of the Poisson noise level ($F_{\rm Poisson}=1$). Interestingly, both the minimum variance and the time at which $\overline{N}=N_T$ are practically independent of the initial mean number $\overline{N_0}$. This enables one to produce the desired number of atoms in a deterministic way without knowing the initial number of atoms. Figure \ref{fig:Histogram} shows the probability distribution of the number of atoms at the time when $\overline{N}=N_T$.

\begin{figure}
\centerline{\scalebox{0.4}{\includegraphics{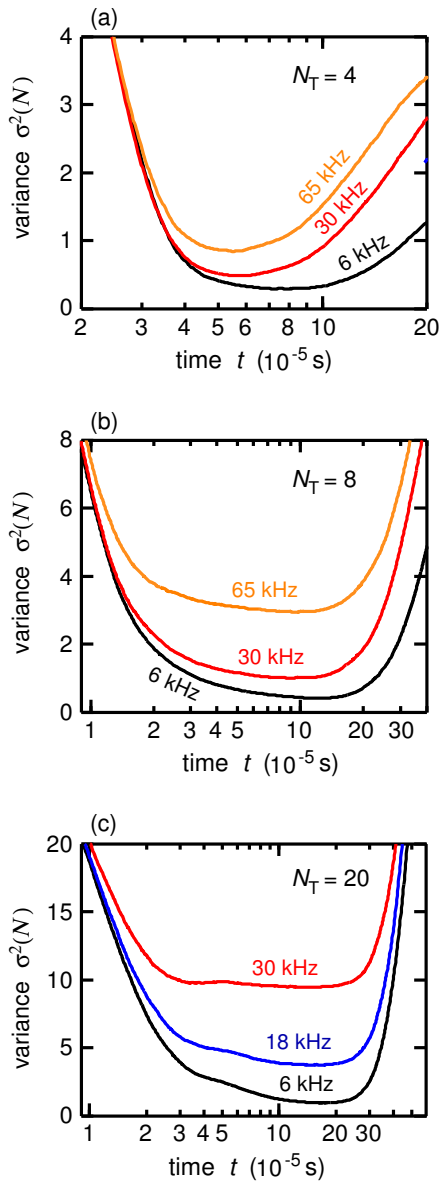}}}
\caption{Effect of the random laser frequency shift for three different target numbers $N_{\rm T}=$4, 8 and 20. (a) Simulations for target number $N_{\rm T}=4$. The interaction energy shift is $\delta = 2 \pi \times 50$ kHz. The variance of the random frequency shift is $\sigma^2(\Delta_{\rm l})/(2 \pi)=$ 6 kHz (black), 30 kHz (red) and 65 kHz (orange). The laser parameters are $\Omega_{\rm c} = 2 \pi \times 2$ MHz, $\Omega_{\rm p} = 2 \pi \times 400$ kHz, and $\Delta_{\rm S} = 2 \pi \times 200$ MHz (from these parameters we obtain $\Delta_{\rm p} \simeq -2 \pi \times 2.9$ MHz and $\Omega_{\rm S} \simeq 2 \pi \times 50$ MHz, using Eqs. \ref{Eq_DeltaP} and \ref{Eq_CharacteristicShift}). (b) Simulations for target number $N_{\rm T}=8$. The interaction energy shift is $\delta = 2 \pi \times 20$ kHz.  The variance of the random frequency shift is $\sigma^2(\Delta_{\rm l})/(2 \pi)=$ 6 kHz (black), 30 kHz (red) and 65 kHz (orange). The laser parameters are $\Omega_{\rm c} = 2 \pi \times 2$ MHz, $\Omega_{\rm p} = 2 \pi \times 400$ kHz, and $\Delta_{\rm S} = 2 \pi \times 300$ MHz ($\Delta_{\rm p} \simeq -2 \pi \times 2.2$ MHz, $\Omega_{\rm S} \simeq 2 \pi \times 54$ MHz). (c) Simulations for target number $N_{\rm T}=20$. The interaction energy shift is $\delta = 2 \pi \times 5$ kHz.  The variance of the random frequency shift is $\sigma^2(\Delta_{\rm l})/(2 \pi)=$ 6 kHz (black), 18 kHz (blue), and 30 kHz (red). The laser parameters are $\Omega_{\rm c} = 2 \pi \times 1$ MHz, $\Omega_{\rm p} = 2 \pi \times 200$ kHz, and $\Delta_{\rm S} = 2 \pi \times 300$ MHz ($\Delta_{\rm p} \simeq -2 \pi \times 1$ MHz, $\Omega_{\rm S} \simeq 2 \pi \times 38$ MHz).} \label{fig:EffectOfNoise}
\end{figure}

We have investigated the impact of an unintended frequency mismatch between the lasers and the atomic transitions caused by possible errors in the experimental system. EIT is very sensitive to two-photon detunings caused by inaccurate frequency locking of any of the two fields \cite{Fleischhauer:05}. To investigate this effect, we have realized stochastic simulations assuming a random frequency shift of the field $\Omega_{\rm p}$, with values of $\sigma^2(\Delta_{\rm l})$ between $2 \pi \times 6$ kHz and $2 \pi \times 65$ kHz (see Eq. \ref{Eq_Hamiltonian}). Figure \ref{fig:EffectOfNoise} shows the variance of the number of atoms $\sigma^2 (N)$ for different cases with target numbers $N_{\rm T}=$ 4, 8 and 20. The minimum of $\sigma^2 (N)$ is reached approximately at the time at which $\overline{N}=N_{\rm T}$. As $\sigma^2(\Delta_{\rm l})$ increases, the atom number fluctuations get larger and less sub-Poissonian. This is because the difference between $\Gamma_{N_{\rm T}}$ and $\Gamma_{N_{\rm T}+1}$ becomes smaller (see Fig. \ref{fig:AtomLossRate}), thus making the minimum of $\Gamma_{N}$ at $N_{\rm T}$ less pronounced. In general, $\sigma^2(\Delta_{\rm l})$ should be smaller than $\delta$ (see Eq. \ref{Eq_CharacteristicShift}) for an effective reduction in atom number fluctuations. The numerical simulations show that Fano factors as low as $F \simeq 0.05$ can be achieved with a good control of the laser frequencies.

The effect of thermal movement is negligible in comparison to the considered effect of random laser frequency mismatch. This is because thermal movement causes the same Doppler shifts in both single-photon transitions, $|g\rangle \leftrightarrow |e\rangle$ and $|q\rangle \leftrightarrow |e\rangle$, thus leaving the energy of the two-photon transition $|g\rangle \leftrightarrow |q\rangle$ unchanged. Since the EIT signal depends on two-photon detunings much more than on single-photon detunings \cite{Fleischhauer:05}, thermal movement at typical ultra cold temperatures can be neglected.

\section{DISCUSSION} \label{sec:disscusion}

This paper has described a method for the generation of strongly sub-Poissonian atom number distributions. The desired number of atoms $N_{\rm T}$ can be achieved in an almost deterministic way. The reduction of atom number fluctuations is accomplished by removing the excess atoms through spontaneous emission into untrapped states. This is an important difference from previous works, in which the removal of atoms was accomplished by making the trap shallower than the interatomic interaction energy of the excess atoms \cite{Campo:08,Pons:09,Sokolovski:11,Chuu:05,Serwane:11,Wenz:13,Whitlock:10,Itah:10,Schlosser:01,Nelson:07,Sortais:12}. Therefore, the control of the trap parameters is less relevant in this paper. On the other hand, this paper requires the precise stabilization of the laser frequency with kilohertz resolution. This can be achieved with state-of-the-art laser technology \cite{Ludlow:07}. Also, coherent ultraviolet light for Rydberg dressing has been experimentally demonstrated with sufficiently high power \cite{Li:07,Bai:19}. Another requirement is that the atoms must be contained within the effective volume of the Rydberg dressing mechanism, whose critical distance is $R = \left| C_6 / (2 \hbar \Delta_{\rm S}) \right|^{1/6}$ \cite{Balewski:14,Gil:14}. For principal quantum numbers $n \simeq 70$, the van-der-Waals coefficients are of the order of $C_6 \simeq 10^{-57}$ Jm$^6$ \cite{Cano:12,Singer:05}, corresponding to critical distances of $\sim 3.5 \mu$m. The required atomic confinement can be provided by tight magnetic microtraps \cite{Boetes:18}.

The advantage of Rydberg dressing is that the interatomic interactions are equally suitable for both fermions and bosons. In general, the Rydberg blockade is a promising resource for atom number control in small atomic ensembles. There are previous works which used resonant excitation into a Rydberg state in order to control $N$. In a pioneering experimental work, atoms were sequentially transferred one-by-one between two ground states, thus creating sub-Poissonian distributions with $\overline{N}=1$ and 2 \cite{Ebert:14}. Another work proposed a method to filter out single Rydberg atoms from ensembles with unknown numbers of atoms \cite{Petrosyan:15}.

To conclude, this paper proposes an efficient solution to the problem of the probabilistic loading of atom traps. The ability to reduce atom number fluctuations and to control the number of atoms in a deterministic way is an enabling toolbox for quantum technologies based on ultracold atoms as well as for fundamental studies of few-body interactions.

\vspace{3mm}
\appendix

\section{AC Stark shift induced by Rydberg dressing} \label{sec:StarkShift}

The AC Stark shift is the interaction energy between the off-resonant laser field $\Omega_{\rm S}$ and the blockaded atomic ensemble. The matrix of the atom-field interaction Hamiltonian in the basis of states $\{ |g^N\rangle, |g^{N-1} r \rangle \}$ is
\begin{equation*} \label{Eq:HS}
{\cal H}_{\rm S} = - \frac{\hbar}{2}
\begin{pmatrix}
0 & \sqrt{N} \Omega_{\rm S}  \\
\sqrt{N} \Omega_{\rm S} & 2 \Delta_{\rm S}
\end{pmatrix},
\end{equation*}
where the factor $\sqrt{N}$ accounts for the collective enhancement of the Rabi frequency. Equation \ref{Eq:StarkShift} is calculated in a trivial way from the series expansion of the low-energy eigenvalue of ${\cal H}_{\rm S}$, as shown in Refs. \cite{Balewski:14,Bouchoule:02}.

\vspace{3mm}

\section{Decay rate from the optical Bloch equations} \label{sec:BlochEquations}

There is an approximate way to estimate the decay rate $\Gamma_N$ by using the optical Bloch equations of a three-level atom in an EIT configuration. The population in the intermediate state $|e\rangle$ of the stationary solution is given by \cite{Gavryusev:16}
\begin{equation}
P_e=\frac{\Omega_{\rm p}}{\Gamma_e} \frac{4 \Delta_{\rm p}^2 \Omega_{\rm p} \Gamma_e}{\left( \Omega_{\rm c}^2 - 4 \Delta_{\rm p}^2 \right)^2 +4 \Delta_{\rm p}^2 \Gamma_e^2}
\label{Eq_Pe_BlochEquations}
\end{equation}
To calculate the population in $|e\rangle$ for $N$ atoms, the right side of Eq. \ref{Eq_Pe_BlochEquations} has to be multiplied by $N$, and the detuning $\Delta_{\rm p}$ has to be substituted by $2 \left( N_{\rm T}-N \right) \delta$, which is the frequency mismatch between the transition $|g^{N}\rangle \leftrightarrow |g^{N-1} e\rangle$ and the laser field $\Omega_{\rm p}$ (see Eqs. \ref{Eq:StarkShift} and \ref{Eq_DeltaP}). The formula for $N$ atoms involves two approximations. First, it neglects the small population in the Rydberg state. Second, it does not consider that the transition $|g^{N}\rangle \leftrightarrow |g^{N-1} e\rangle$ has an energy different than that of the transition $|g^{N-1}e\rangle \leftrightarrow |g^{N-2} e^2\rangle$, and therefore the collective state cannot be expressed as product of individual atomic states. Nevertheless, the formula reproduces the same tendency as the numerical simulations with the multi-atom Hamiltonian of Eq. \ref{Eq_Hamiltonian}, as we can see in Fig. \ref{fig:AtomLossRate}. Although only the numerical simulations with the multi-atom Hamiltonian are precise, the approximate formula provides better understanding of the EIT origin of $\Gamma_N$ as a function of $N$.

\end{document}